# ENHANCED MOTOR IMAGERY-BASED EEG CLASSIFICATION USING A DISCRIMINATIVE GRAPH FOURIER SUBSPACE

*Maliheh Miri[1], Vahid Abootalebi[1], Hamid Behjat[2]*

[1] Department of Electrical Engineering, Yazd University, Yazd, Iran
[2] Department of Biomedical Engineering, Lund University, Lund, Sweden

## ABSTRACT

Dealing with irregular domains, graph signal processing (GSP) has attracted much attention especially in brain imaging analysis. Motor imagery tasks are extensively utilized in brain-computer interface (BCI) systems that perform classification using features extracted from Electroencephalogram signals. In this paper, a GSP-based approach is presented for two-class motor imagery tasks classification. The proposed method exploits simultaneous diagonalization of two matrices that quantify the covariance structure of graph spectral representation of data from each class, providing a discriminative subspace where distinctive features are extracted from the data. The performance of the proposed method was evaluated on Dataset IVa from BCI Competition III. Experimental results show that the proposed method outperforms two state-of-the-art alternative methods.

***Index Terms***— graph signal processing, EEG, simultaneous diagonalization, classification.

## 1. INTRODUCTION

Electroencephalogram (EEG) is a non-invasive, high temporal resolution brain imaging modality that captures functional and physiological changes within the brain [1, 2]. Motor imagery (MI) tasks are dynamic states during which neuronal activity in the primary sensorimotor areas modifies similar to a real executed movement [3]. MI tasks are extensively utilized in brain-computer interface (BCI) systems and can be classified by extracting features from EEG signals to identify a user's mental state [4, 5]. Many methods have been proposed to classify MI tasks from EEG signals. Some approaches are based on extracting key information from the time and frequency domains [6-8]. Some other approaches of MI classification are focused on learning spatial filters from multichannel EEG signals to extract discriminative features from data [9, 10]. There are also many studies which have proposed applying mathematical transforms, such as wavelet transforms, to extract discriminative features via decomposition of EEG signals [11, 12].

The recently emerged field of graph signal processing (GSP) [13-15] has attracted great interest in different signal processing applications, in particular, signals defined on irregular domains such as the human brain [16-18]. In [19] the role of the GSP on the classification and dimensionality reduction of functional MRI (fMRI) data was evaluated. Promising results have also been presented that suggest the benefits of GSP in classification of EEG signals [20, 21].

In this work, using EEG data, we define a brain graph that characterizes the temporal correlation structure between the EEG electrodes. We then transform the EEG data into a spectral graph representation. The covariance structure of the resulting spectral representations is then computed, resulting in a matrix for each class of data. A classification framework is then proposed, in which simultaneous diagonalization of these two matrices provides the basis of a discriminative subspace that can be used to differentiate the two motor imagery tasks. An exploratory analysis is then performed to identify which spectral graph components from the data provide the most discriminative features. Results from the proposed method are also compared to two alternative methods that use simultaneous diagonalization.

The remainder of this paper is structured as follows. Section 2 gives an overview of the fundamental concepts and the proposed framework. Section 3 presents the experimental results and provides a discussion. Section 4 presents our concluding remarks.

## 2. MATERIALS AND METHODS

### 2.1. Graph signal processing fundamentals

Let $\mathcal{G} = (\mathcal{V}, \mathcal{E}, \mathbf{A})$ denote an undirected, weighted graph, where $\mathcal{V} = \{1, 2, \ldots, N\}$ denotes the graph's finite set of N vertices, $\mathcal{E}$ denotes the graph's edge set, i.e., pairs $(i, j)$ where $i, j \in \mathcal{V}$, and $\mathbf{A}$ is a symmetric matrix that denotes the graph's weighted adjacency matrix. To exploit the spectral properties of the graph, the graph's normalized Laplacian matrix is defined as $\mathcal{L} = \mathbf{I} - \mathbf{D}^{-1/2}\mathbf{A}\mathbf{D}^{-1/2}$, where $\mathbf{D}$ is the diagonal matrix of vertex degrees, i.e., $\mathbf{D}_{ii} = \sum_j \mathbf{A}_{ij}$ and $\mathbf{I}$ is the identity matrix. Let $\ell^2(\mathcal{G})$ denote the Hilbert space of all square-integrable graph signals $\mathbf{f}: \mathcal{V} \rightarrow \mathbb{R}$ that are defined on $\mathcal{V}$;

a graph signal $\mathbf{f} \in \ell^2(\mathcal{G})$ is an N×1 vector, whose $n$-th component represents the signal value at the $n$-th vertex of $\mathcal{G}$. Since $\mathcal{L}$ is a real and positive semi-definite matrix, it can be diagonalized via eigenvalue decomposition as:

$$\mathcal{L} = \mathbf{U}\mathbf{\Lambda}\mathbf{U}^T, \quad (1)$$

where $\mathbf{U} = \{\mathbf{u}_1, \mathbf{u}_2, ..., \mathbf{u}_N\}$ is a matrix of orthonormal eigenvectors of $\mathcal{L}$ (an eigenvector in each column) with corresponding eigenvalues $0 = \lambda_1 \leq \lambda_2 \leq ... \leq \lambda_N \leq 2$ in the diagonal matrix $\mathbf{\Lambda}$. The eigenvalues define the graph Laplacian spectrum, and the corresponding eigenvectors form an orthonormal basis that spans the $\ell^2(\mathcal{G})$ space. By using the Laplacian eigenvectors, a graph signal $\mathbf{f}$ can be transformed into a spectral representation, commonly denoted as the graph Fourier transform (GFT) of $\mathbf{f}$, obtained as $\hat{\mathbf{f}} = \mathbf{U}^T\mathbf{f}$; the inverse GFT is obtained as $\mathbf{f} = \mathbf{U}\hat{\mathbf{f}}$. Importantly, the GFT satisfies Parseval's energy conservation relation [22], i.e., $\|\mathbf{f}\|_2^2 = \|\hat{\mathbf{f}}\|_2^2$. Graph Laplacian eigenvectors corresponding to larger eigenvalues entail a larger extent of variability, and as such, eigenvalues of the graph Laplacian matrix can be seen as an extension of frequency elements that define the Fourier domain in classical signal processing [15].

**2.2. Data description**

In order to evaluate the proposed method, we used EEG signals from the publicly available BCI Competition III-Dataset IVa [23]. The signals were recorded from five healthy subjects (labelled as *aa*, *al*, *av*, *aw*, and *ay*) using 118 electrodes arranged in the extended international 10/20-system at a sampling rate of 100 Hz. Subjects were presented with 280 3.5-second-long visual cues during which they were asked to perform right hand or right foot motor imageries; 140 trials were acquired for each class. According to the competition instructions, for each class the trials were divided into training and test sets, wherein the set sizes differed across the five subjects. The first two subjects have the most labelled trials (60% and 80%, respectively), while the other three have 30%, 20% and 10% labelled trials, respectively; as such, performing classification is more challenging on subjects av, aw, and ay due to their small training set size.

**2.3. Graph-based representation of brain signals**

We modeled the structure of the brain by a graph with vertices corresponding to the EEG electrodes and edges quantifying the degree of functional connectivity between the electrodes in each subject. Let $\mathbf{f}_{i,t}$ and $\mathbf{f}_{j,t}$ denote the time series of electrodes $i$ and $j$, respectively. The absolute value of the Pearson correlation between $\mathbf{f}_{i,t}$ and $\mathbf{f}_{j,t}$, providing an estimation of statistical dependency of the two temporal signals, was considered as the weight of the edge connecting vertices $i$ and $j$.

For each trial, we used the time points within the 0.5-2.5 second interval after the visual cue to construct graph signal; this 2-second interval has been previously proposed by the winner of BCI Competition III-Dataset IVa. Given that motor activity, be it real or imagined, causes modulations of the mu and beta rhythms [5], we filtered the extracted signal with a third-order Butterworth filter with a pass band of 8-30 Hz. Graph signals were then extracted from these filtered signals; in particular, we defined one graph signal per time instance, i.e., a signal representing EEG values across the 118 electrodes, which thus resulted in T=200 graph signals per trial. We then used the eigenvectors of the EEG graph normalized Laplacian matrix to compute the GFT of each signal. As such, we obtain a representation of brain signals that jointly encodes structural, functional, and temporal characteristics of the data.

**2.4. Discriminative subspace through simultaneous diagonalization**

Inspired by the methods presented in [18, 24], simultaneous diagonalization of two matrices was considered to provide a discriminative subspace for two-class (right hand and right foot) MI classification. For graph signal $\mathbf{f}$ defined on $\mathcal{G}$, let $\tilde{\mathbf{f}}$ denote the de-meaned and normalized version of $\mathbf{f}$, obtained as [16]:

$$\tilde{\mathbf{f}} = \frac{(\mathbf{f} - \mathbf{u}_1^T \mathbf{f} \mathbf{u}_1)}{\|(\mathbf{f} - \mathbf{u}_1^T \mathbf{f} \mathbf{u}_1)\|_2}. \quad (2)$$

More precisely, let $\mathbf{F}_k$ denote an N×T matrix with elements $\{f_{c,t}\}$, where $c = 1, ..., N$ and $t = 1, ..., T$, denote the $k$-th trial of the EEG time series, where c denotes electrode number; similarly, let $\hat{\tilde{\mathbf{F}}}_k$ denote the GFT matrix of the $k$-th de-meaned and normalized trial. The goal is to determine a transform $\hat{\mathbf{P}}$ that simultaneously diagonalizes the following two symmetric matrices that are computed for each class:

$$\bar{\Xi}_1 = \frac{1}{K_1} \sum_{k=1}^{K_1} \frac{\hat{\tilde{\mathbf{F}}}_{1k} \hat{\tilde{\mathbf{F}}}_{1k}^T}{\text{tr}(\hat{\tilde{\mathbf{F}}}_{1k} \hat{\tilde{\mathbf{F}}}_{1k}^T)}$$
$$\bar{\Xi}_2 = \frac{1}{K_2} \sum_{k=1}^{K_2} \frac{\hat{\tilde{\mathbf{F}}}_{2k} \hat{\tilde{\mathbf{F}}}_{2k}^T}{\text{tr}(\hat{\tilde{\mathbf{F}}}_{2k} \hat{\tilde{\mathbf{F}}}_{2k}^T)}, \quad (3)$$

where $T$ and tr(.) denote the transpose and the trace operator, respectively, and $K_i$ is the number of the trials in class *i*. As a first step, we whiten $\bar{\Xi} = \bar{\Xi}_1 + \bar{\Xi}_2$ such that:

$$\mathbf{P}^T \bar{\Xi} \mathbf{P} = \mathbf{P}^T (\bar{\Xi}_1 + \bar{\Xi}_2) \mathbf{P} = \dot{\Xi}_1 + \dot{\Xi}_2 = \mathbf{I}. \quad (4)$$

Due to positive definiteness of $\bar{\Xi}$, whitening transform $\mathbf{P}$ can be derived via singular value decomposition as:

$$\bar{\Xi} = \Phi \Theta \Phi^T; \quad \mathbf{P} = \Phi \Theta^{-\frac{1}{2}}. \quad (5)$$

Consequently, eigenvalue decomposition of $\dot{\Xi}_1$ gives:

$$\dot{\Xi}_1 = \Psi \Theta_1 \Psi^T \longrightarrow \dot{\Xi}_2 = \Psi(\mathbf{I} - \Theta_1)\Psi^T. \quad (6)$$

In particular, $\dot{\Xi}_1$ and $\dot{\Xi}_2$ share the same eigenvectors but their eigenvalues are complementary; that is, the eigenvector associated with the largest eigenvalue of $\dot{\Xi}_1$ corresponds to the smallest eigenvalue of $\dot{\Xi}_2$. Therefore, a small combination of the first and last eigenvectors of $\Psi$ induces a suitable discriminatory transform for differentiating the two classes. Finally, the overall transformation matrix can be obtained as:

$$\hat{\mathbf{P}} = \Psi^T \mathbf{P}. \quad (7)$$

This matrix was used to project the GFT coefficients of the de-meaned and normalized graph signals to a discriminative feature space; these features were then used for classification.

## 3. RESULTS AND DISCUSSION

In our experiments, the algorithms were trained using the training set data available for each subject, and consequently, the test set data available for each subject were used to evaluate the performance of the methods by assigning a label to each trial. The variance of the projected GFT coefficients on the first and the last rows of $\hat{\mathbf{P}}$ were used to train an SVM classifier with a linear kernel. This projection maximizes the variance of the signals from one class while minimizing it for the signals of the other class [24].

In the first experiment, the GFT coefficients were used in four different settings. In the first setting, we used the entire set of GFT coefficients, i.e., all frequencies (AF), whereas in the second to fourth settings we only used a subset of the coefficients by equally dividing them into three sub-bands, low (LF), medium (MF) and high (HF) frequencies, respectively; division of the spectrum into 3 sub-bands was inspired by prior work on application of GFT on brain imaging data [25, 26]. These four sets of GFT coefficients were then used to derive the discriminative matrix $\hat{\mathbf{P}}$, and consequently, features for classification were extracted by projecting them on $\hat{\mathbf{P}}$. Table 1 shows the classification accuracies for each individual subject and also on average across subjects.

**Table 1.** Classification accuracy (%) on the test sets in four different frequency band settings.

|    | *aa*  | *al*  | *av*  | *aw*   | *ay*  | Mean ± std    |
|----|-------|-------|-------|--------|-------|---------------|
| AF | 69.64 | **100** | 71.43 | 92.41  | 81.75 | 83.05 ±13.15 |
| LF | **87.50** | **100** | 70.92 | **94.19** | **90.08** | **88.54 ±10.92** |
| MF | 52.68 | 76.78 | 55.10 | 54.012 | 54.36 | 58.59 ± 10.21 |
| HF | 57.14 | 69.64 | 49.49 | 56.69  | 48.81 | 56.36 ± 8.39  |

In all subjects, classification accuracy obtained by using the LF GFT coefficients substantially outperforms that resulting from using the MF and HF coefficients, and moreover, it outperforms using all the GFT coefficients on average across subjects as well as in subjects aa, aw, and ay. There seems to be no correlation between the rise in performance and the training set size of the subjects.

Considering the substantial classification accuracy obtained by using the lowest one-third of the GFT coefficients, we attempted to find the optimal subset of the graph frequencies that present the most discriminative features for classification. To this end, we performed 10-fold cross-validation (CV) on the training sets across all the frequencies in each subject. Fig. 1 shows the mean accuracies achieved for each subject across all eigenvalues (frequency elements). The eigenvalue index that resulted in the highest accuracy was chosen as the *subject-specific* (SS) cut-off frequency for specifying the LF GFT coefficients; the resulting SS cut-off frequencies notably varied across subjects, with values in the range [20,117]. Classification results obtained by using the determined SS low frequency bands are presented in Table 2, denoted by SS-LF. Despite these SS frequency bands manifesting the best performance on the training sets in the CV framework, their performance was not generalizable to the test sets, which can be seen as potential overfitting of the selected LF bands to the training sets.

**Table 2.** Classification accuracies on the test sets for the proposed methods and two state-of-the-art methods. Numbers indicated in parenthesis denote the selected cut-off frequencies for selecting the lower end of the spectral band.

|              | *aa*           | *al*    | *av*           | *aw*           | *ay*           | Mean±std      |
|--------------|----------------|---------|----------------|----------------|----------------|---------------|
| SS-LF        | 85.71 (60)     | **100** (47) | 70.41 (117)    | 93.3 (28)      | 88.09 (20)     | 87.50±11.01   |
| O1-LF (23)   | 87.5           | 98.21   | 59.18          | 85.27          | 86.11          | 83.25±14.43   |
| O2-LF (30)   | 89.28          | **100** | 68.37          | 93.3           | 91.27          | 88.44±11.93   |
| O3-LF (32)   | **91.96**      | **100** | 68.88          | 94.2           | **92.46**      | **89.5±11.96** |
| O4-LF (35)   | 89.28          | **100** | 68.88          | 93.30          | 92.06          | 88.71±11.76   |
| SRCSP [9]    | 72.32          | 96.43   | 60.2           | 77.68          | 86.51          | 78.63±13.78   |
| RCSSP [10]   | 82.14          | 96.42   | 68.87          | **98.21**      | 88.88          | 86.91±11.94   |

Given that using SS cut-off frequencies did not result in better performance, we attempted to find an optimal cut-off that can be used across subjects. This was done by performing 10-fold CV on the training sets of each subject for different cut-off frequencies. Fig. 1 shows the resulting classification accuracies across subjects. Overall, on average across subjects, best performance was obtained at a cutoff frequency corresponding to spectral elements within the range [20,40], after which point the performance almost saturated in all five subjects. Four cut-off values within the optimal window were then selected (see vertical dashed lines), the results for which are reported in Table 2, denoted by O$x$-LF, where $x = 1,...,4$

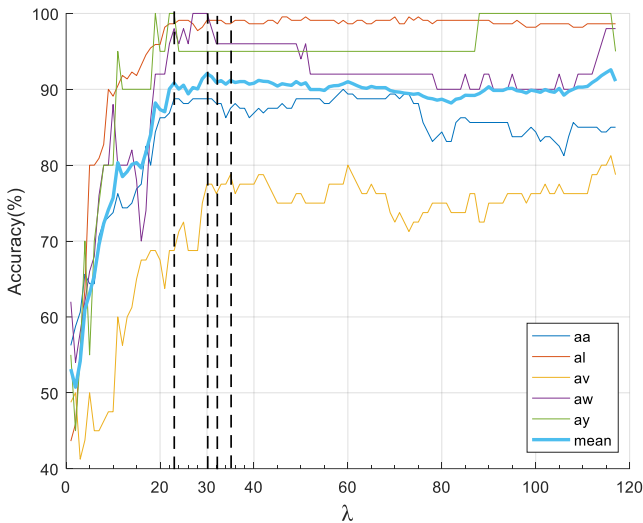

**Fig. 1.** Classification accuracies for 10-fold CV on the training sets for different indices of cut-off frequencies (λ).

Overall, the best average accuracy was obtained by exploiting only the first 32 graph frequencies, which shows a gain of 6.4% compared to utilizing the entire set of GFT elements across the graph spectra (compare AF and O3-LF in Tables 1 and 2, respectively). Furthermore, in comparison with the subject-specific band selection scheme, using O3-LF resulted in approximately 2% higher accuracy than SS-LF on average across subjects, as well as higher accuracies in four out of five subjects.

Overall, these results show that simultaneous diagonalization using the LF GFT coefficients provides a more discriminative subspace compared to using the GFT coefficients across the entire spectrum in all subjects, except for subject av; in subject av the EEG graph signal energy profiles are more broadly spread across the spectrum (results not shown), therefore, suggesting the potential benefit of using a SS definition of the LF band for this subject.

In Table 2. results from two state-of-the-art methods that exploit a similar diagonalization approach as proposed in this paper are presented for comparison. Importantly, in contrast to the proposed approach, these two methods do not take into account the structural information of the brain signals, as defined by the proposed EEG graph. Comparison of the results verifies that combining structure, function, and temporality of the brain signals enhances the classification performance of EEG signals compared to these prior approaches. Furthermore, extracting features from a low dimensional discriminative subspace precludes overfitting to the training sets, which is an important aspect to be considered especially in subjects that have small training sets. Given that the eigenvalues of the Laplacian matrix are considered as the basis of the graph frequency domain, graph frequency analysis encodes the spatial variation of the signals. Lower graph frequency components represent smooth signals that vary slowly across the brain connectivity graph and high graph frequency components represent signals that manifest a greater extent of spatial variation relative to the underlying brain connectivity network [25]. Our experimental results reflect better classification accuracies of the MI tasks by projecting the GFT coefficients on the lower part of the graph spectrum. These observations indicate that imagined motor activities are not localized and represent a regular and smooth pattern across the whole brain. Furthermore, EEG graph signal spectral energies, as reflected by the GFT coefficients, are more concentrated on low spatial frequencies, which to an extent explains the enhanced discriminative performance that was obtained by using only spectral energy content of data at the lower end of the spectra, corroborating similar results that have been observed on fMRI graph signals [25, 27]. We defer detailed investigation of the energy profiles of EEG graph signals to future work.

## 4. CONCLUSIONS

In this paper, a GSP framework was presented for classification of MI tasks from multi-electrode EEG data. Experimental results suggested that decomposition of brain signals on a discriminative subspace of the graph Fourier domain outperforms two related conventional methods for classification of EEG signals. Graph frequency analysis of EEG data on brain functional connectivity graphs reveals the spatially smooth nature of motor activity signals, making it possible to classify these signals by exploiting the representation of the data on the low frequency range of the graph spectrum. In future work, the proposed method will be validated on a dataset with a greater number of subjects. Moreover, our future work will be focused on investigating the intrinsic properties of representing EEG data in the graph Fourier domain, with the goal of finding a generalizable approach to extract discriminative subspaces for other classes of EEG signals. We will also explore alternative methods for deriving brain graphs from EEG data using graph learning techniques [28].

## 5. COMPLIANCE WITH ETHICAL STANDARDS

The present research study was conducted retrospectively using human subject data made available in open access by provided by the Berlin BCI group [23]. Ethical approval for analyzing the openly available data is not required according to our local ethics committee.

## 6. ACKNOWLEDGMENTS

Hamid Behjat was supported by the Swedish Research Council (2018-06689) and in part by the Royal Physiographic Society of Lund. The authors certify that they have no conflict of interest to report in regards to the subject matter discussed in this paper.